\documentclass[aps,12pt,pra,showpacs,superscriptaddress,preprint,a4paper]{revtex4}
\usepackage{ae}
\usepackage[T1]{fontenc}
\usepackage[latin1]{inputenc}
\usepackage{amsmath}
\usepackage{subfig}
\usepackage{array}
\usepackage{amsfonts}
\usepackage{epsf}
\usepackage{epsfig}
\usepackage{amssymb}
\usepackage{bbm}
\usepackage{color}
\begin{document}
\title{Solitary and Jacobi elliptic wave solutions of the generalized Benjamin-Bona-Mahony equation}
\author{\small Didier Belobo Belobo $\surd\surd$}
\email[E-mail: ]{belobodidier@gmail.com}\affiliation{Laboratory
of Nuclear, Atomic, Molecular Physics and Biophysics, University
of Yaounde I} \affiliation{Centre d'Excellence Africain en
Technologies de l'Information et de la Communication (CETIC),
University of Yaounde I} \affiliation{Laboratoire d'Analyses, de
Simulations et d'Essais (LASE), IUT, University of Ngaoundéré}
\affiliation{African Institute for Mathematical Sciences, 6
Melrose Road, Muizenberg, Cape Town, 7945, South Africa}
\author{\small Tapas Das $\surd\surd$}
\email[E-mail: ]{tapasd20@gmail.com}\affiliation{Kodalia Prasanna
Banga High School (H.S), South 24 Parganas, 700146, India}
\begin{abstract}
Exact bright, dark, antikink solitary waves and Jacobi elliptic
function solutions of the generalized Benjamin-Bona-Mahony
equation with arbitrary power-law nonlinearity will be
constructed in this work. The method used to carry out the
integration is the F-expansion method. Solutions obtained have
fractional and integer negative or positive power-law
nonlinearities. These solutions have many free parameters such
that they may be used to simulate many experimental situations,
and to precisely control
the dynamics of the system.\\
Keywords: Generalized Benjamin-Bona-Mahony equation, F-expansion
method, solitary waves, Jacobi elliptic function solutions.
\end{abstract}
\pacs{02.30.Jr, 05.45.Yv,  47.35.Fg} \maketitle
\section*{1. Introduction}

The search of exact solutions for evolution nonlinear partial
differential equations (NLPDEs) plays an important role in the
study of nonlinear physical phenomena. This is due to the fact
that, nonlinear phenomena are ubiquitous in nature, thus appear
in a wide range of fields in physics, mathematical physics, and
engineering. Some celebrated evolution NLPDEs are the
Hasegawa-Mima equation [1] which describes turbulence in plasma
physics, the Fitzhugh-Nagumo equation [2] that models biological
neuron, the Hunter-Saxton equation [3-4] used to study waves
orientation in nematic liquid crystal, the nonlinear
Schr\"{o}dinger equation that models the dynamics of waves in
many media such as matter waves in Bose-Einstein condensates [5],
the evolution of electromagnetic fields in fiber optics [6], the
evolution of gravity driven surface water waves [7], the
evolution of the order parameter in the BCS theory [8], just to
name a few.

In the past five decades, a great deal of attention has been paid
to the dynamics of shallow water waves, mainly modeled by an
evolution NLPDE known as the Korteweg-de Vries (KdV) equation
[9], and modified KdV equations [9]. The KdV equation is valid
when the water depth is constant and is derived under the
assumption of small wave-amplitude and large wave length.
Modified KdV equations include KdV equations with varying bottom
and higher order corrections of the KdV equation. Solutions of
the KdV and modified KdV equations have actively been
investigated, and include solitary waves which come from a
delicate balance between dispersion and nonlinearity, periodic
waves like the Jacobi elliptic function solutions and so on
[9,10].

In 1972, the regularized long-wave equation, better known as the
the Benjamin-Bona-Mahony (BBM) equation [11] were introduced as a
regularized form of the KdV equation. As pointed out in [11,12],
the BBM equation better describes long waves and, as far as the
existence, uniqueness and stability are concerned, the BBM
equation has some substantial advantages over the KdV equation
[11]. Moreover, the BBM equation also finds applications in other
contexts such as the modeling of the drift of waves in plasma
physics or the Rossby waves in rotating fluids [13], wave
transmission in semi-conductors and optical devices [14],
hydromagnetic waves in cold plasma, acoustic waves in anharmonic
crystals, and acoustic gravity waves in compressible fluids [11].
In the past four decades, the BBM equation and its various
versions have been intensively studied, and many types of
solutions have been found among them solitary waves and periodic
waves which can be found in the explicit literature. A
classification of some forms of the BBM equation and their
solutions can be found in [15]. Unlike the KdV equation, the BBM
equations are not exactly integrable in the sense of the Painlevé
test of integrability. Nevertheless, various techniques have been
developed which help to carry out the integration of BBM
equations. Among them are the tanh and the sine-cosine methods
[16], the Jacobi elliptic function expansion method [17], the
first integral method [18], the variable-coefficient
balancing-act method [19], the hyperbolic auxiliary function
method [20], the homogeneous balance method [21]. A generalized
(1+1) BBM equation with dual arbitrary power-law nonlinearity may
be written in dimensionless form as [22,23]
\begin{eqnarray}
u_t+\alpha u_x+(\beta u^n+\gamma u^{2n})u_x-\delta u_{xxt}=0,
\end{eqnarray}
where $u$ is the wave profile, $\alpha$ and $\delta$ are the
dispersion coefficients, $n$ is the arbitrary power-law
nonlinearity, $\beta$ and $\gamma$ the coefficients of the dual
power-law nonlinearity. For $\alpha=1$, $\beta=1$, $n=1$,
$\gamma=0$, and $\delta=1$ Eq.(1) recovers the BBM equation.
Wazwa\textcolor[rgb]{0.00,0.00,1.00}{z} solved Eq.(1) in the case
where $\alpha=1$, $\delta=-1$, $\beta=0$ by means of the tanh and
the sine-cosine methods, and obtained solitary and periodic
solutions [24]. Yang, Tang, and Qiao constructed solitary and
periodic wave solutions of Eq.(1) for $\alpha=0$, $n>0$ and
$\delta\neq 0$ using an improved tanh function method [22]. Liu,
Tian, and Wu used the Weierstrass elliptic function method to
construct two solutions of Eq.(1) for $\alpha=0$ in terms of the
Weierstrass elliptic functions [25]. Biswas employed the solitary
wave ansatz method and proposed a one-soliton solution of Eq.(1)
[23]. All the latter works show the importance of investigating
solutions of the \textcolor[rgb]{0.00,0.00,1.00}{BBM equation
given by Eq.(1)}. However, to the best of our knowledge, Eq.(1)
with non-vanishing coefficients has only been tackled in the work
of Ref. [23] where a one-soliton solution was found. As an
evolution NLPDE with important applications in different fields
in physics, it is important to find more solutions of Eq.(1) that
may help to have a better understanding of physical phenomena or
at least give orientations for future applications. For example,
solitary and Jacobian elliptic function solutions have been
intensively used for practical applications in physics and
engineering; these solutions have not been fully investigated for
the generalized BBM equation
\textcolor[rgb]{0.00,0.00,1.00}{(Eq.(1))} with all non-vanishing
coefficients.

The aim of this work is to construct solitary and Jacobi elliptic
wave solutions of the generalized BBM equation
\textcolor[rgb]{0.00,0.00,1.00}{(Eq.(1))} with all non-vanishing
coefficients. To this end, we use the F-expansion method
introduced in [26] which has been an accurate tool to integrate
evolution NLPDEs, along with the auxiliary ordinary equation
[5,27]. The paper is organized as follows, in Sec. 2, we
construct analytical solutions of Eq.(1). Then we discuss the
characteristics and evolution of the solutions in Sec. 3. The
paper is concluded in Sec. 4.
\section*{2. Analytical solutions}

We start our quest of analytical solutions of Eq.(1) by setting
the following traveling wave transformation
\begin{eqnarray}
u(x,t)=U(\zeta)\,,\, \zeta =kx-Vt \,,
\end{eqnarray}
where $k$ is the inverse of the width of the wave and $V$ its
velocity. Inserting Eq.(2) into Eq.(1) we obtain a nonlinear
ordinary differential equation for the function $U$
\begin{eqnarray}
(\alpha k-V)U_{\zeta} + k(\beta U^{n}+\gamma
U^{2n})U_{\zeta}+\delta k^{2}VU_{\zeta\zeta\zeta}=0\,,
\end{eqnarray}
with $U_{\zeta}\equiv\frac{\partial U}{\partial\zeta}$ and
$U_{\zeta\zeta\zeta}\equiv\frac{\partial^{3}
U}{\partial\zeta^{3}}$. An integration of Eq.(3) yields
\begin{eqnarray}
(\alpha k-V)U + k(\frac{\beta}{n+1} U^{n+1}+\frac{\gamma}{2n+1}
U^{2n+1}) + \delta k^{2}VU_{\zeta\zeta}=0\,,
\end{eqnarray}
where the right-hand side constant of integration has been set to
zero. Eq.(4) is difficult to solve analytically, in order to find
analytical solutions, we need to transform it into a more
tractable and manageable form. Toward that end, we use the
transformation $\omega = U^{n}$. After a little algebra, a
nonlinear ordinary differential equation in terms of the function
$\omega$ is retrieved
\begin{eqnarray}
\omega\omega_{\zeta\zeta}+p\omega^{2}+q\omega^{3}+r\omega^{4}+s{(\omega_{\zeta})}^2=0\,,
\end{eqnarray}
in which the parameters $p$, $q$, $r$, $s$ are given by
\begin{subequations}
\begin{align}
p&=\frac{(\alpha k-V)n}{\delta k^{2}V}\,,\\
q&=\frac{\beta n}{\delta k(n+1)V}\,,\\
r&=\frac{\gamma n}{\delta k(2n+1)V}\,,\\
s&=\frac{1-n}{n}\,,
\end{align}
\end{subequations}
with these forbidden values for $n$: $-1, -1/2, 0, 1$.

The next step is to solve Eq.(5) by means of the F-expansion
method. Thus, following the standard procedure [5,26,27], we seek
the solution in the form
\begin{eqnarray}
\omega(\zeta)=\sum_{i=0}^{N}a_iF^i(\zeta)\,,
\end{eqnarray}
where the function $F$ satisfies the auxiliary equation
\begin{eqnarray}
\Big(\frac{dF}{d\zeta}\Big)^2=b_0+b_1F(\zeta)+b_2F^2(\zeta)+b_3F^3(\zeta)+b_4F^4(\zeta)\,.
\end{eqnarray}
Exact solutions of the auxiliary equation (8) can be found in
Ref. [27], while the coefficients
$a_{i}$\textcolor[rgb]{0.00,0.00,1.00}{,} $b_{j}$ ($j=0,1,2,3,4$)
will be determined later. The value of the integer $N$ is found
by balancing the highest nonlinear and derivative terms. This
idea in our case leads to $N=1$. Hence,
\begin{equation}
\omega(\zeta)=a_0+a_1F(\zeta).
\end{equation}
Inserting Eq.(9) along with Eq.(8) into Eq.(5), and collecting
all the coefficients of powers of $F$ ($F^{m}(\zeta)$, $m = 0, 1,
2, 3, 4$), setting each coefficient to zero, yields a set of
algebraic equations for the unknowns $a_{0}$, $a_{1}$, $q$, $r$,
$b_{j}$ ($j=0,1,2,3,4$) and or $s$
\begin{subequations}
\begin{align}
a_0a_1b_1+2pa_0^2+2qa_0^3+2ra_0^4+2sa_1^2b_0&=0\,,\\
2a_0a_1b_2+a_1^2b_1+4pa_0a_1+6qa_0^2a_1+8ra_0^3a_1+2sa_1^2b_1&=0\,,\\
2a_1^2b_2+3a_0a_1b_3+2pa_1^2+6qa_0a_1^2+12ra_0^2a_1^2+2sa_1^2b_2&=0\,,\\
4a_0a_1b_4+3a_1^2b_3+2qa_1^3+8ra_0a_1^3+2sa_1^2b_3&=0\,,\\
2a_1^2b_4+ra_1^4+sa_1^2b_4&=0.
\end{align}
\end{subequations}
As explained in [27], the solutions of the auxiliary equation are
sensitive to specific values of the coefficients $b_j$ ($j = 0,
1, 2, 3, 4$) and can be regrouped into two families: family (I)
$b_0 = b_1 = 0$; family (II) $b_1 = b_3 = 0$. We show below that
these two families lead to solitary waves and Jacobi elliptic
function solutions of Eq.(1).
\subsection*{2.1 Family I solutions: $b_0 = b_1 = 0$}

We simplify the system \textcolor[rgb]{0.00,0.00,1.00}{of
Eqs.(10)} by setting for example $p=-1$ and obtain the following
solutions:
\begin{subequations}
\begin{align}
a_{01}&= 0, a_{11}=a_{11}, b_{21}=\frac{1}{s+1},
q_{1}=-\frac{b_3(2s+3)}{2a_{11}}, b_{41}=-\frac{a_{11}^2r}{s+2}\,,\\
a_{02}&=\sqrt{\frac{-(s+2)}{r(s+1)}}, a_{12}=\frac{b_3}{2}a_{02},
b_{22}=b_{21},
q_2=\frac{(2s+3)}{2a_{02}}, b_{42}=\frac{-b_3^2(s+1)}{4}\,,\\
a_{03}&=-a_{02}, a_{13}=-a_{12}, b_{23}=b_{21}, q_3=-q_2,
b_{43}=b_{42}.
\end{align}
\end{subequations}
Therefore, it is possible to obtain solitary wave solutions if we
consider these solutions of the auxiliary equation (8):
\begin{subequations}
\begin{align}
b_0&=b_1=0, b_2>0, F_{I,1}(\zeta) =
\frac{-b_{2}b_{3}\textrm{sech}^{2}(\sqrt{b_{2}}/2)\zeta)}{
b^{2}_{3}-b_{2}b_{4}[1-\tanh(\sqrt{b_{2}/4}\zeta)]^{2}}\,,\\
b_0&=b_1=0, b_2>0, b_3^2-b_2b_4>0, F_{I,2} =
\frac{2b_{2}\textrm{sech}(\sqrt{b_{2}}\zeta)}{\sqrt{b^{2}_{3}-4b_{2}b_{4}}-
b_3\textrm{sech}(\sqrt{b_{2}}\zeta)}.
\end{align}
\end{subequations}
Hence, the analytical solution of Eq.(1) is written as
\begin{equation}\label{13}
u_{Im}(x,t) = (a_{0m'} + a_{1m'}F_{Im}(\zeta))^{\frac{1}{n}},
\end{equation}
where $m = 1, 2$ and $m' = 1, 2, 3$. Solutions
\textcolor[rgb]{0.00,0.00,1.00}{given by Eq.(13)} are correct
ones if that of the set (11) are compatible with the requirements
of the parameters $b_j$ of the set
\textcolor[rgb]{0.00,0.00,1.00}{of Eq.(12)}. These compatibility
conditions impose some restrictions to our solutions Eq.(13). To
be precise, in the set of Eqs.(12), $b_2>0$ implies $n>0$. This
means that our solutions correspond to only positive values of
the power-law nonlinearities (in addition to $n\neq 1$ obtained
above). When $m=2$ in Eq.(13), the condition $b_3^2-4b_2b_4>0$
imposes $\mid b_3\mid>2n\mid a_1\mid\sqrt{\frac{-r}{n+1}}$;
without loss of generality, we consider that $k, V >0$ and taking
into account the expression of $r$ given by Eq.(6c), $r$ must be
negative, hence $\gamma$ and $\delta$ have opposite signs.
Moreover, other important information can be extracted from the
set of Eqs.(11). On the first hand, after an examination of
Eq.(11a) (zero background solutions $a_{01}=0$), one deduces that
$a_{1}$, $b_3$, $\gamma$, $\delta$, $k$, and $V$ are free
parameters, while $\alpha = \frac{V(1-\frac{\delta k^2}{n})}{k}$
and $\beta = \frac{-b_3kV(n+2)(n+1)\delta}{2n^2a_{1}}$. The
experimenter has many possibilities to manipulate some physical
parameters of the solution like its width, velocity, amplitude,
and to consider different physical situations by playing with the
values and signs of $\gamma$, $\delta$, and $n$, meanwhile
altering the signs and values of $\alpha$, $\beta$. For $m=1$ in
Eq.(13), solitary wave solutions are obtained when $\delta$ and
$\gamma$ have the same signs, a situation consistent with the
existence of solitary waves in many physical system due to a
balance between dispersion and nonlinearity [5,6] (and references
therein). Additional specific conditions lead to bright or dark
solitary waves. A bright solitary wave is obtained in the
following cases: (i) $a_1$ and $b_3$ with opposite signs and $n$
is an integer, a sample is presented in Fig. 1(a) for $n=2, m' =
1$; (ii) $n=1/n'$, $n'$ even, an example is depicted in Fig. 1(b)
where $n' = 2, m' = 1$; (iii) $a_1$, $b_3$ with opposite signs
and $n=1/n'$, $n'$ odd, Fig. 1(c) ($n' = 3, m' = 1$) displays
such a case. A black solitary wave solution is obtained if $a_1$
and $b_3$ have opposite signs and $n = 1/n'$ with $n'$ odd as can
be seen in Fig. 1(d) ($n' = 3, m' = 1$). On the second hand, we
observe that Eqs.(11b) and (c) may give rise to solitary wave
solutions on finite backgrounds since $a_{0}$ is real and does
not vanish, $\beta =
\pm\frac{kV\delta(n+2)(n+1)}{2n^2}\sqrt{\frac{-r}{n}}$. For $m=1$
in Eq.(13), compatible conditions on $r$ and $b_2$ are identical
to the case of Eq.(11a), while for $m=2$, the compatible
condition $b_3^2 - 4b_2b_4>0$ is always satisfied ($b_3^2 -
4b_2b_4 = 2b_3^2$). We now present solitary waves embedded on
finite backgrounds. To this end, we focus on the solution
\textcolor[rgb]{0.00,0.00,1.00}{Eq.(13)} with $m = 2$ and recall
that $\delta$ and $\gamma$ must have opposite signs. Bright
solitary waves are obtained if $b_3>0$, $n = 1/n'$ and (i) $n'$
even (see Fig. 2(a) $n' = 2, m' = 3$), (ii) $n'$ odd with $a_0,
a_1>0$ (see Fig. 2(b) $n' = 3, m' = 3$). A dark solitary wave is
obtained with the conditions (ii) but $a_0, a_1<0$ (see Fig. 2(c)
$n' = 3, m' = 2$). For $b_3<0$, Eq.(13) represents a dark
solitary wave in many cases, but in some cases where the
background is sufficiently small and negative, $n'$ being an odd
integer, an 'antibright' solitary wave solution can be obtained
(Fig. 2(d) $n' = 3, m' = 2$).
\subsection*{2.2 Family II solutions $b_0 = b_1 = 0$}

In this part of the work, we show that Eq.(1) also admits Jacobi
elliptic function solutions. The solutions of the set
\textcolor[rgb]{0.00,0.00,1.00}{of Eqs.(10)} for $b_1 = b_3 = 0$
are
\begin{subequations}
\begin{align}
b_{01}&=\frac{a^{2}_{04}(p+4a^{2}_{04}r)}{4a^{2}_{14}},
b_{24}=2a^{2}_{04}r-\frac{p}{2},
q_{4}=\frac{-4a^{2}_{04}r+p}{2a_{04}}\,,\\
b_{44}&=\frac{a^{2}_{14}(p-4a^{2}_{04}r)}{a^{2}_{04}},
s=\frac{-2(p-2a^{2}_{04}r)}{p-4a^{2}_{04}r}.
\end{align}
\end{subequations}
Taking advantage of four Jacobi elliptic function solutions of
the auxiliary \textcolor[rgb]{0.00,0.00,1.00}{Eq.(8)}:
\begin{subequations}
\begin{align}
b_{01}&=\frac{1}{4}, b_{24}=\frac{k_1^2-2}{2}\,,\\
F_{II1}(\zeta)&=\frac{\textrm{cn}(\zeta,
k_{1})}{\sqrt{1-k_1^2}+\textrm{dn}(\zeta, k_{1})},
F_{II2}(\zeta)=\frac{\textrm{cn}(\zeta,
k_{1})}{\sqrt{1-k_1^2}-\textrm{dn}(\zeta,
k_{1})}\,,\\
F_{II3}(\zeta)&=\frac{\textrm{sn}(\zeta,
k_{1})}{1+\textrm{dn}(\zeta, k_{1})},
F_{II4}(\zeta)=\frac{\textrm{sn}(\zeta,
k_{1})}{1-\textrm{dn}(\zeta, k_{1})},
\end{align}
\end{subequations}
the Jacobi elliptic function solutions of Eq.(1) are
\begin{equation}\label{16}
u_{IIm}(x,t) = (a_{0m'} + a_{1m'}F_{IIm}(\zeta))^{\frac{1}{n}},
\end{equation}
where $m = 1, 2, 3, 4$, $k_1$ ($0<k_1<1$) denotes the modulus of
Jacobi elliptic functions. As already explained above, correct
solutions of Eq.(16) require that the parameters $b_j$ in
\textcolor[rgb]{0.00,0.00,1.00}{Eq.(15a)} be compatible with
their counterparts of the set \textcolor[rgb]{0.00,0.00,1.00}{of
Eqs.(14)}, providing the expressions of $a_0, a_1, p, k_1$
\begin{subequations}
\begin{align}
a_{04}&=\sqrt{\frac{-(n+1)}{3nr}},
a_{14}=\frac{1}{3}\sqrt{\frac{-(n+1)}{nr}},
p=\frac{-4}{3n}, k_1=\frac{\sqrt{6}}{3}\,,\\
a_{05}&=a_{04}, a_{15}=-a_{14}, p, k_1\,,\\
a_{06}&=-a_{04}, a_{16}=a_{14}, p, k_1\,,\\
a_{07}&=-a_{04}, a_{17}=-a_{14}, p, k_1.
\end{align}
\end{subequations}
It is simple to deduce from the set of
\textcolor[rgb]{0.00,0.00,1.00}{Eqs.(17) that in Eq.(16)} $m' =
4, 5, 6, 7$, $\alpha = \frac{V}{k}(\frac{-4\delta k^2}{3n}+1)$,
and $\beta = \frac{\delta k(n+1)Vq_4}{n}$, while $q_4$ is given
in \textcolor[rgb]{0.00,0.00,1.00}{Eq.(14)}. The parameters $n,
\gamma, \delta, k, V$ are free in solutions
\textcolor[rgb]{0.00,0.00,1.00}{Eq.(16)}. In order to represent
some periodic solutions, we suppose that $k, V, n >0$,
consequently, $\gamma$ and $\delta$ have opposite signs since $r$
should be negative in the set \textcolor[rgb]{0.00,0.00,1.00}{of
Eqs.(17)}. Furthermore, the modulus $k_1 = \sqrt{6}/3$ is always
constant. We plot in Fig 3 two periodic solutions of Eq.(1); in
panel (a) $n = 2, m = 1, m' = 7$, while in panel (b) $n = 2, m =
3, m' = 7$.
\section*{3. Discussion}

In [23] first appeared a bright solitary wave solution of Eq.(1)
with all non null \textcolor[rgb]{0.00,0.00,1.00}{coefficients,}
the arbitrary power-law nonlinearity $n$ belonging to the
interval $0<n<2$. In \textcolor[rgb]{0.00,0.00,1.00}{Sec. 2}, we
have proposed bright, black, dark, and Jacobi elliptic function
solutions of \textcolor[rgb]{0.00,0.00,1.00}{Eq.(1)} with all
non-vanishing coefficients, and show that the power-law
nonlinearity $n$ is allowed to take more positive values than in
[23] ($n>0, n\neq 1$). Hence, the solutions presented in
\textcolor[rgb]{0.00,0.00,1.00}{Sec. 2} may be used to probe the
validity of the model (Eq.(1)) with higher power-law
nonlinearities. Moreover, the two families of solutions also
admit negative values of the power-law nonlinearity $n$. For
example, Jacobi elliptic function
\textcolor[rgb]{0.00,0.00,1.00}{solutions given by Eq.(16)} allow
negative $n$ in the following cases: (i) $-1<n<0$, $r>0$ which
means that $\delta, \gamma$ have the same signs; (ii) $n<-1$,
$r<0$ thus $\delta, \gamma$ have opposite signs. We display in
Fig. 4(a) \textcolor[rgb]{0.00,0.00,1.00}{the solution expressed
by Eq.(16)} for $n = -3, m = 3, m' = 4$. Besides,
\textcolor[rgb]{0.00,0.00,1.00}{the solution Eq.(13)} for
negative $n$ holds only for $m = 1$ (the case $m = 2$ does not
satisfy the condition $b_3^2-4b_2b_4>0$). Setting $p = 1$, the
solutions of the set \textcolor[rgb]{0.00,0.00,1.00}{of Eqs.(10)}
with $n<0$ become $a_{01}= 0, a_{11}=a_{11},
b_{21}=\frac{-1}{s+1}, q_{1}=-\frac{b_3(2s+3)}{2a_{11}},
b_{41}=-\frac{a_{11}^2r}{s+2}$,
$a_{02}=\sqrt{\frac{(s+2)}{r(s+1)}}, a_{12}=\frac{-b_3}{2}a_{02},
b_{22}=b_{21}, q_2=\frac{-(2s+3)}{2a_{02}}$,
$b_{42}=\frac{-b_3^2(s+1)}{4}, a_{03}=-a_{02}, a_{13}=-a_{12},
b_{23}=b_{21}, q_3=-q_2, b_{43}=b_{42}$. One may expect that
\textcolor[rgb]{0.00,0.00,1.00}{the solution Eq.(13)} gives rise
to rather different solitary wave solutions since the physics of
the model has completely changed. Such an expectation is
confirmed by Fig. 4(b) where an antikink profile solution is
presented for $n = -1/3, m = 1, m' = 2$.

\section*{4. Conclusion}

This paper obtains bright, dark, antikink, 'antibright' solitary
waves and Jacobi elliptic function solutions of the generalized
BBM equation with dual power-law nonlinearity. The F-expansion
method (along with the auxiliary equation) is used to integrate
the BBM equation. Solutions are constructed for arbitrary
power-law nonlinearities that might be positive or negative,
fractional or integer. Many physical parameters may be used to
mimic a wide range of experimental scenarios, whereas the
solutions constructed may be used to probe the accuracy of the
model with higher power-law nonlinearities. The issue of
stability of the solutions of the generalized BBM equation needs
a thorough treatment that shall be addressed by means of
perturbation techniques and numerical simulations in future
works.

\section*{FIGURE CAPTIONS}
\subsection*{Figure 1}
(Color online) Spatiotemporal evolution of solitary wave
solutions with positive power-law nonlinearity given by Eq.(13).
(a) $n = 2$, $\alpha = -1$, $\beta = 0.1$, $\gamma = 2$, $\delta
= 2$, $V = 1$, $k = 1$, $b_{3} = 0.1$, $a_{1} = -1$. (b) Same
parameters as in panel (a) except $n = 1/2$ and $a_{1} = 1$. (c)
Same parameters as in panel (a) except $n = 1/3$, $\beta = 0.28$.
(d) Same parameters as in panel (b) except $n = 1/3$, $\beta =
-0.28$. (a)-(c) bright solitary waves, (d) black solitary wave.
\subsection*{Figure 2}
(Color online) (a)-(b) bright, (c) dark solitary wave solutions
(13) on finite backgrounds. (a) $n = 1/2$, $\alpha = 2$, $\delta
= -1$, $\beta = 5\sqrt{3}$, $a_{0} = \sqrt{3}$, $a_{1} =
0.05\sqrt{3}$, $m' = 3$. (b) $n = 1/3$, $\beta = 2.8\sqrt{30}$,
$a_{0} = \sqrt{30}/3$, $a_{1} = 0.01666\sqrt{30}$. (c) $n = 1/3$,
$\beta = -0.28\sqrt{30}$, $a_{0} = -\sqrt{30}/3$, $a_{1} =
-0.01666\sqrt{30}$, $m' = 2$. Other parameters in (a)-(c) are the
same as in Fig. 1(a). (d) $a_{1} = 0.01666\sqrt{30}$, other
parameters as in (c).
\subsection*{Figure 3}
(Color online) Spatiotemporal evolution of Jacobi elliptic
function solutions (16). (a) $m = 1$, (b) $m = 3$. Parameters are
$n = 2$, $\alpha = 2/3$, $\beta = 4\sqrt{5}/5$, $\gamma = -1$,
$\delta = 1$, $a_{0} = \sqrt{5}/2$, $a_{1} = \sqrt{15}/6$, $m' =
7$. Other parameters as in Fig 1(a).
\subsection*{Figure 4}
(Color online) Solutions with negative power-law nonlinearity.
(a) Jacobi elliptic function solution (16), $n = -3$, $\alpha =
23/27$, $\beta = \frac{2}{45}\sqrt{30}$, $a_{0} = \sqrt{30}/9$,
$a_{1} = \sqrt{10}/9$, $m' = 4$, other parameters as in Fig.
3(b). (b) Antikink solution (13), $n = -1/3$, $\alpha = 4$,
$\beta = 10\sqrt{3}$, $\delta = -1$, $\gamma = 2$, $a_{0} =
-\sqrt{3}/3$, $a_{1} = 0.0166\sqrt{3}$, $m = 2$, $m' = 1$.
\begin{figure}[!ht]
 \begin{center}
  \includegraphics[width=5cm, height=5 cm]{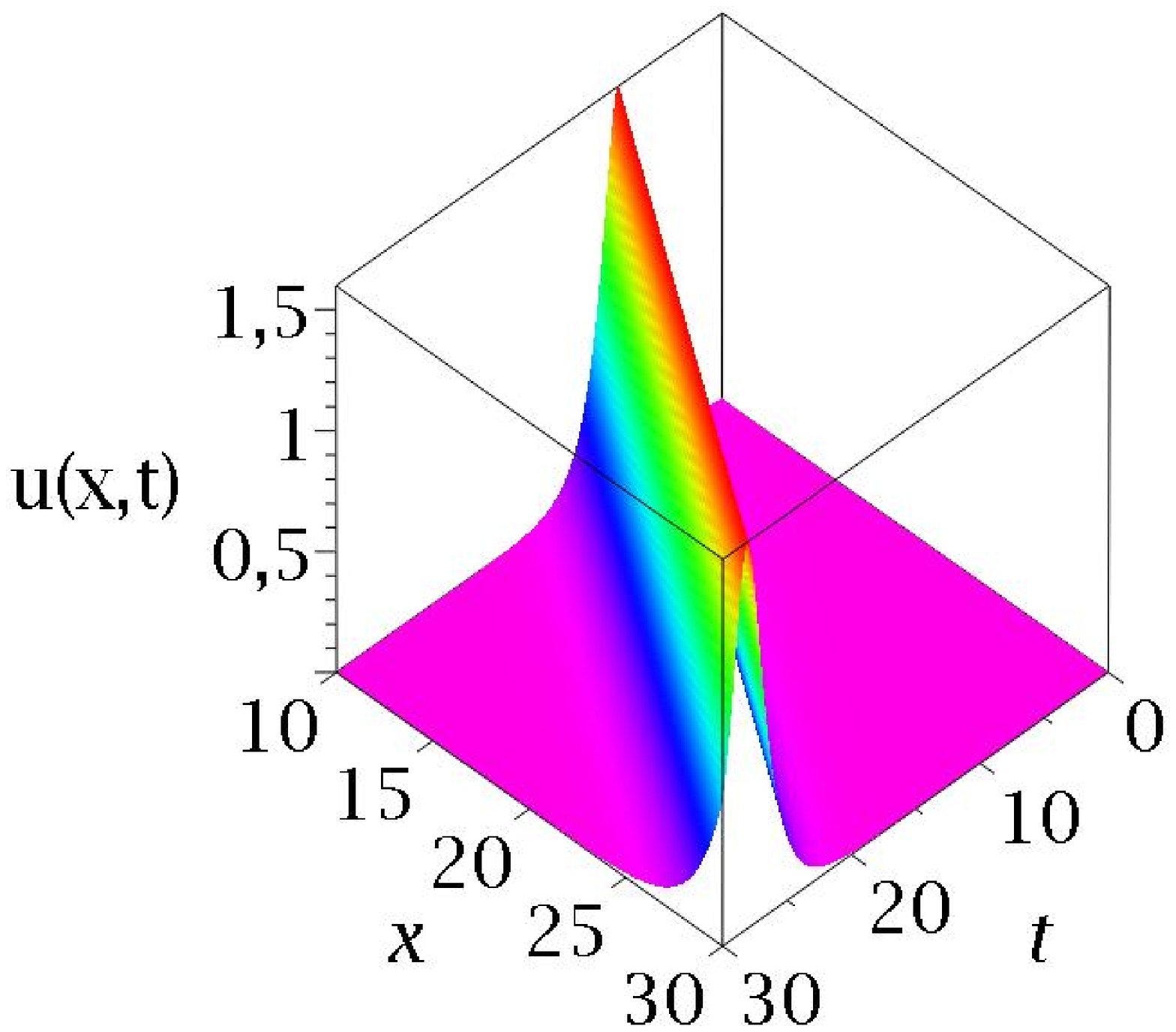}(a)
  \includegraphics[width=5cm, height=5 cm]{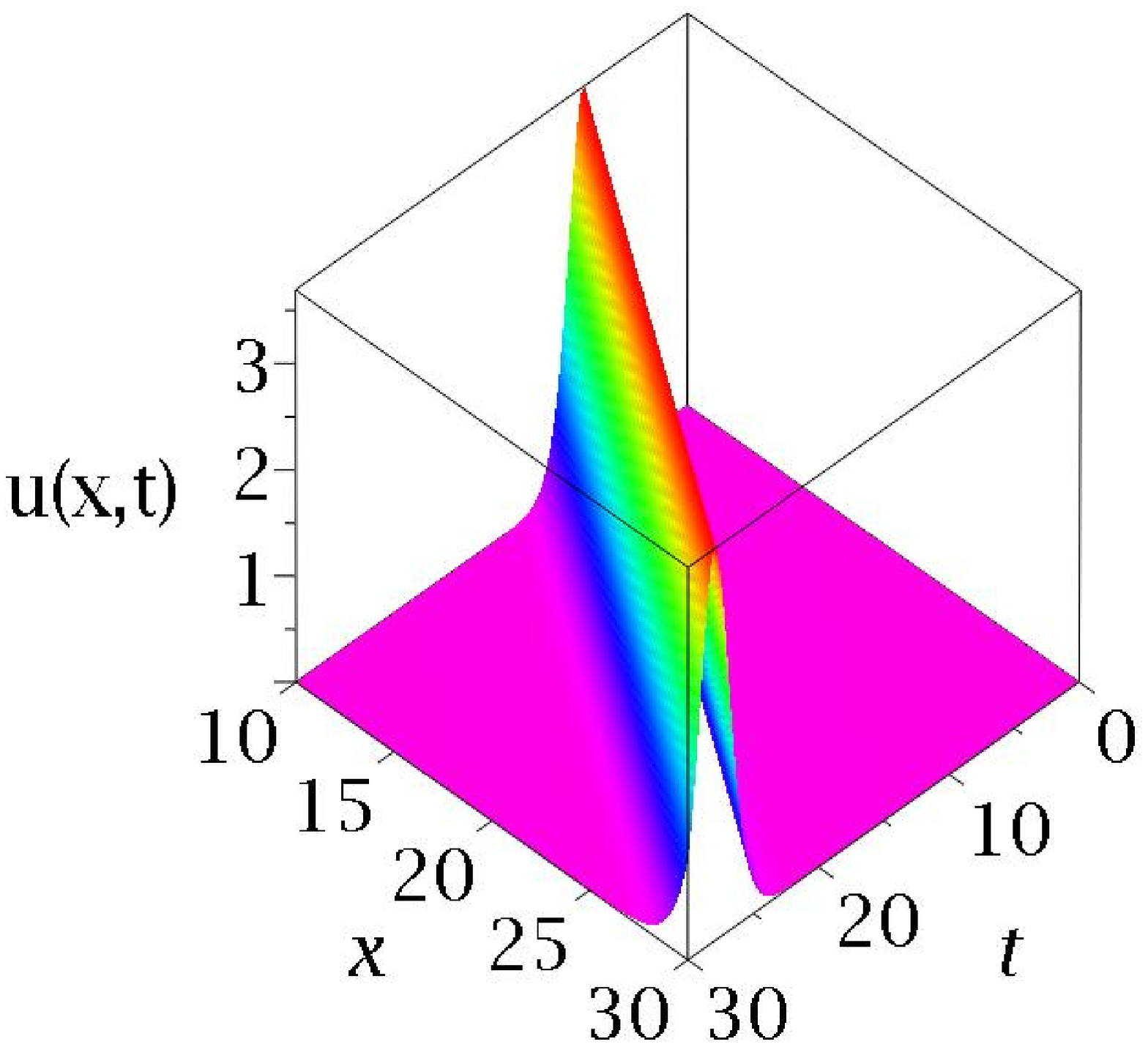}(b)
  \includegraphics[width=5cm, height=5 cm]{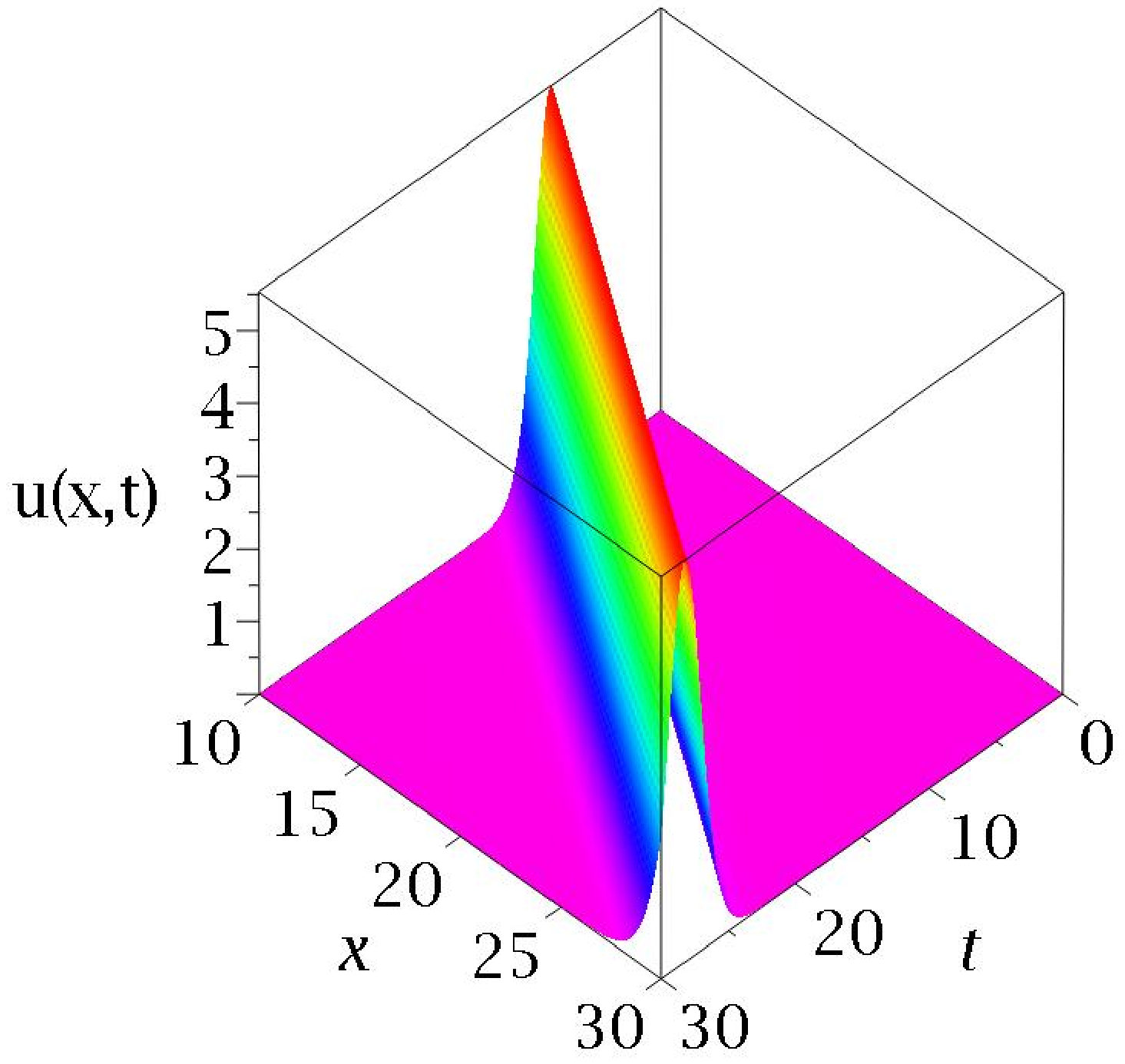}(c)
  \includegraphics[width=5cm, height=5 cm]{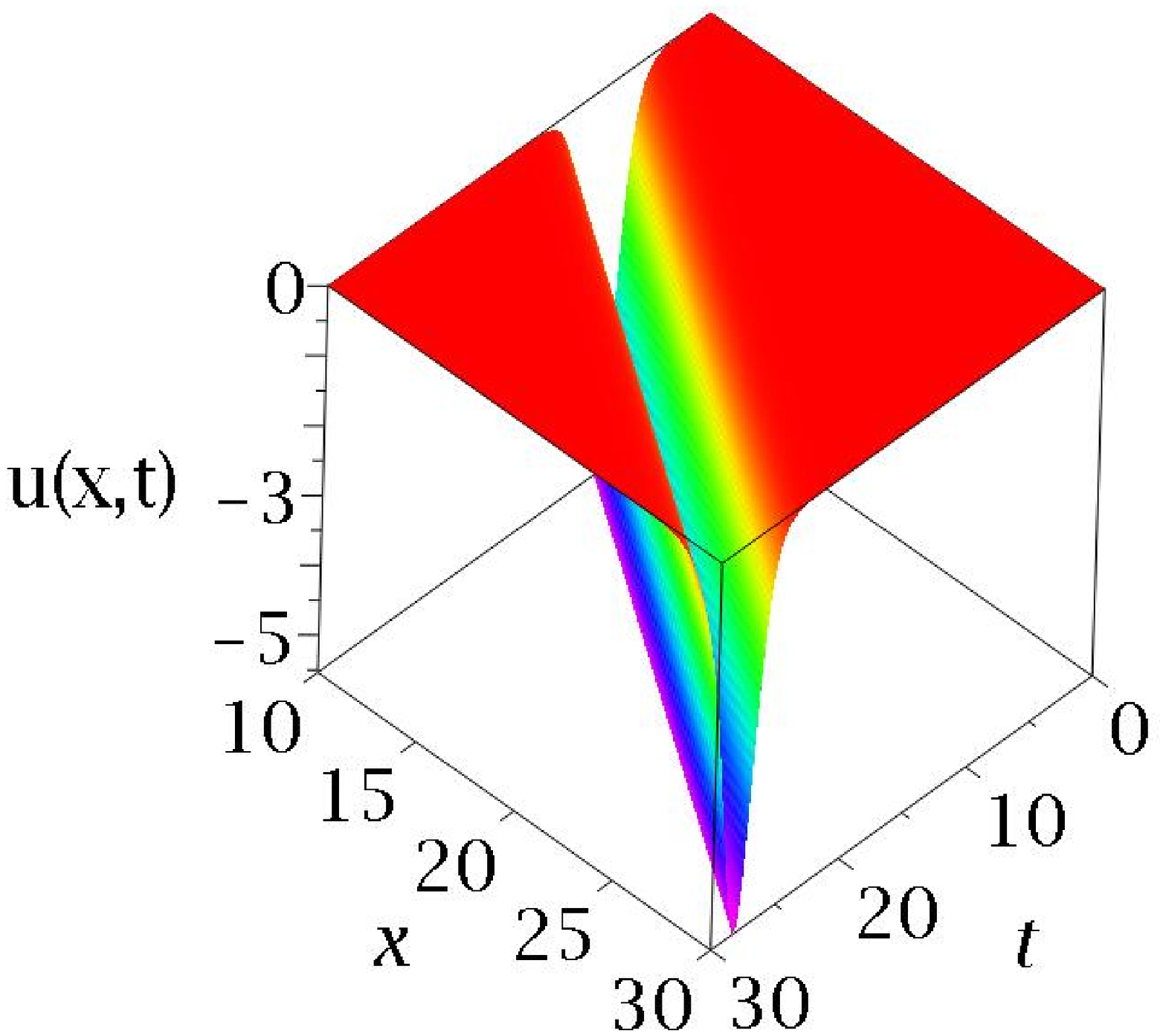}(d)
    \caption{}\label{1}
 \end{center}
\end{figure}

\begin{figure}[!ht]
 \begin{center}
  \includegraphics[width=5cm, height=5 cm]{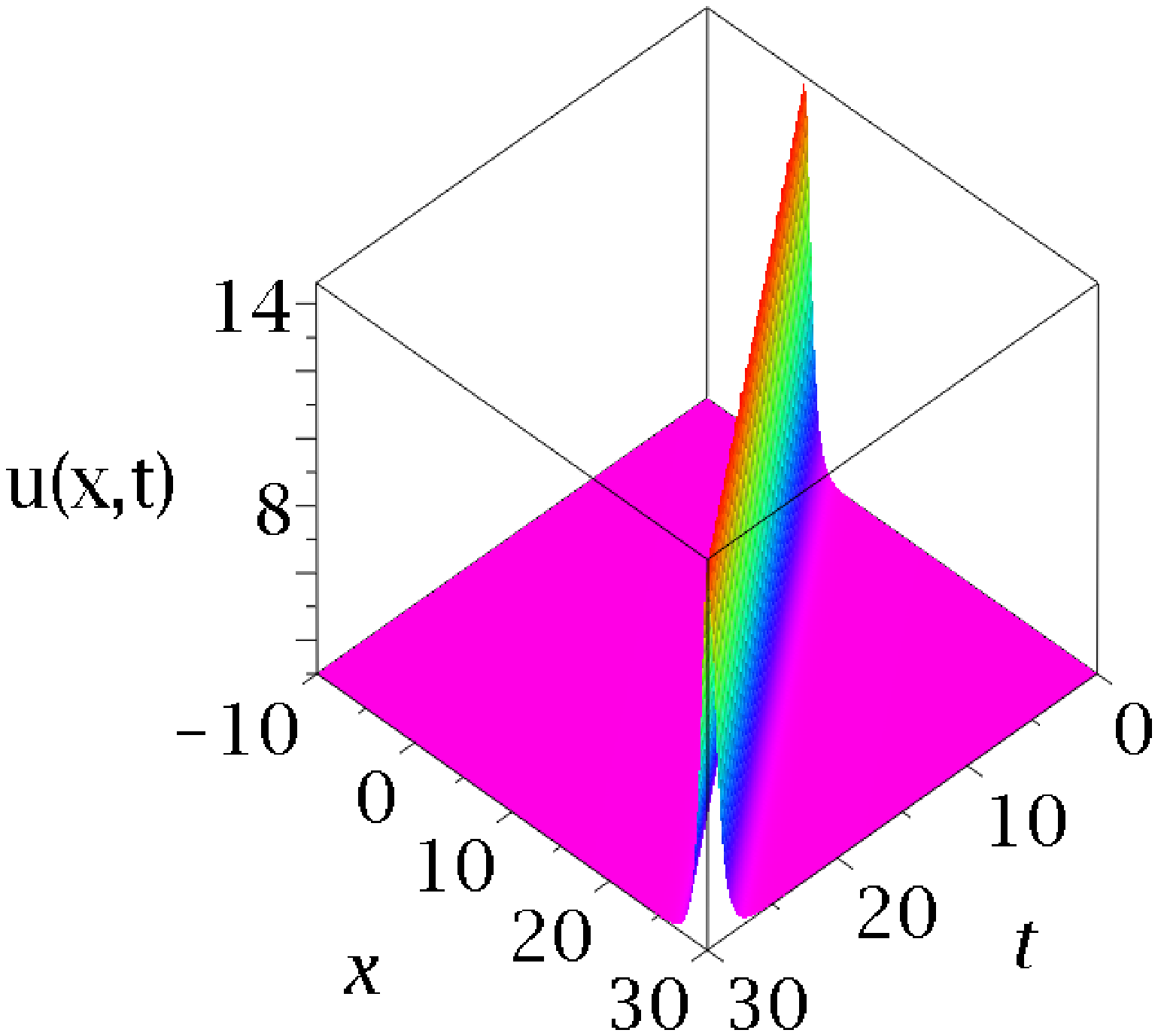}(a)
  \includegraphics[width=5cm, height=5 cm]{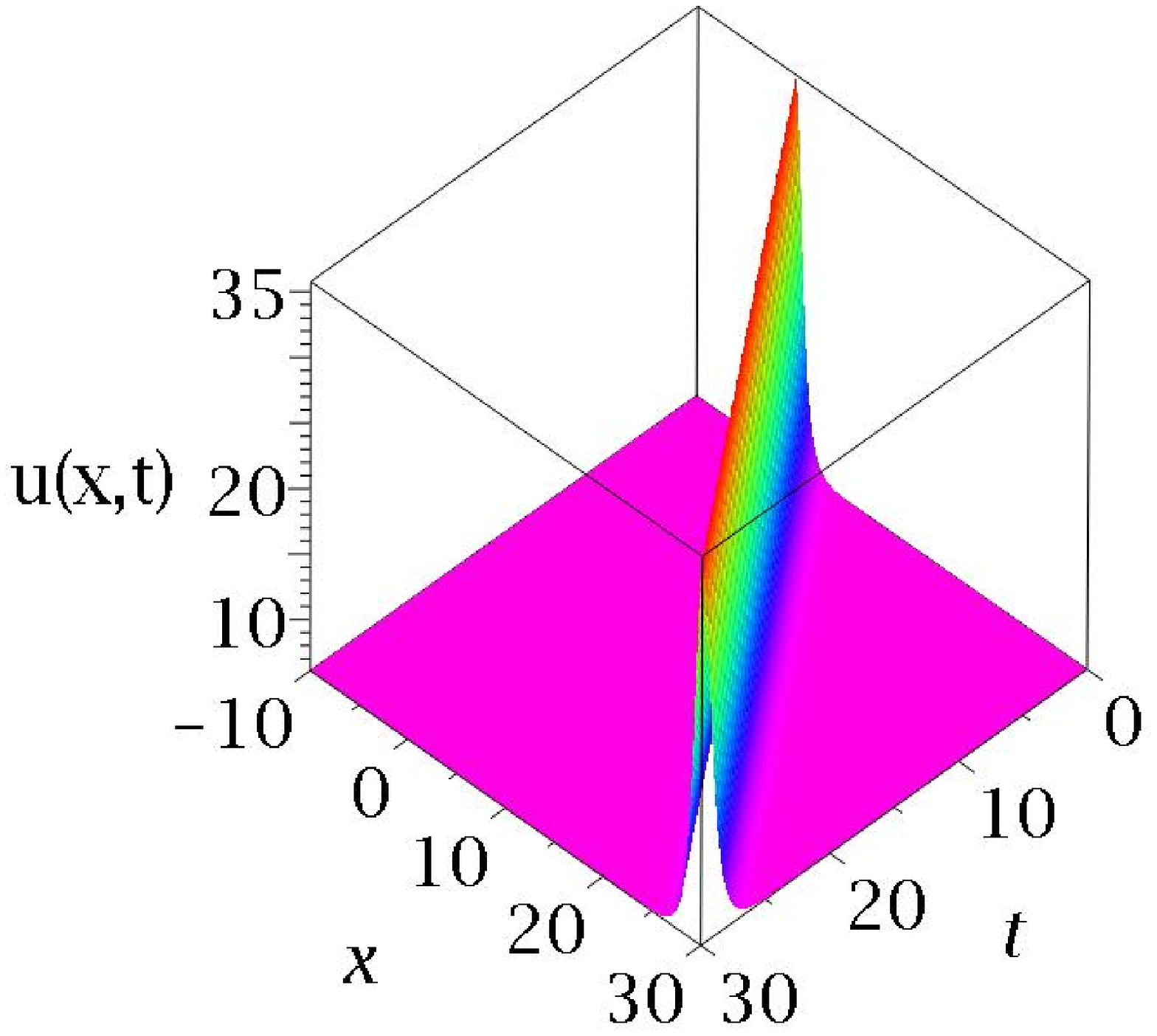}(b)
  \includegraphics[width=5cm, height=5 cm]{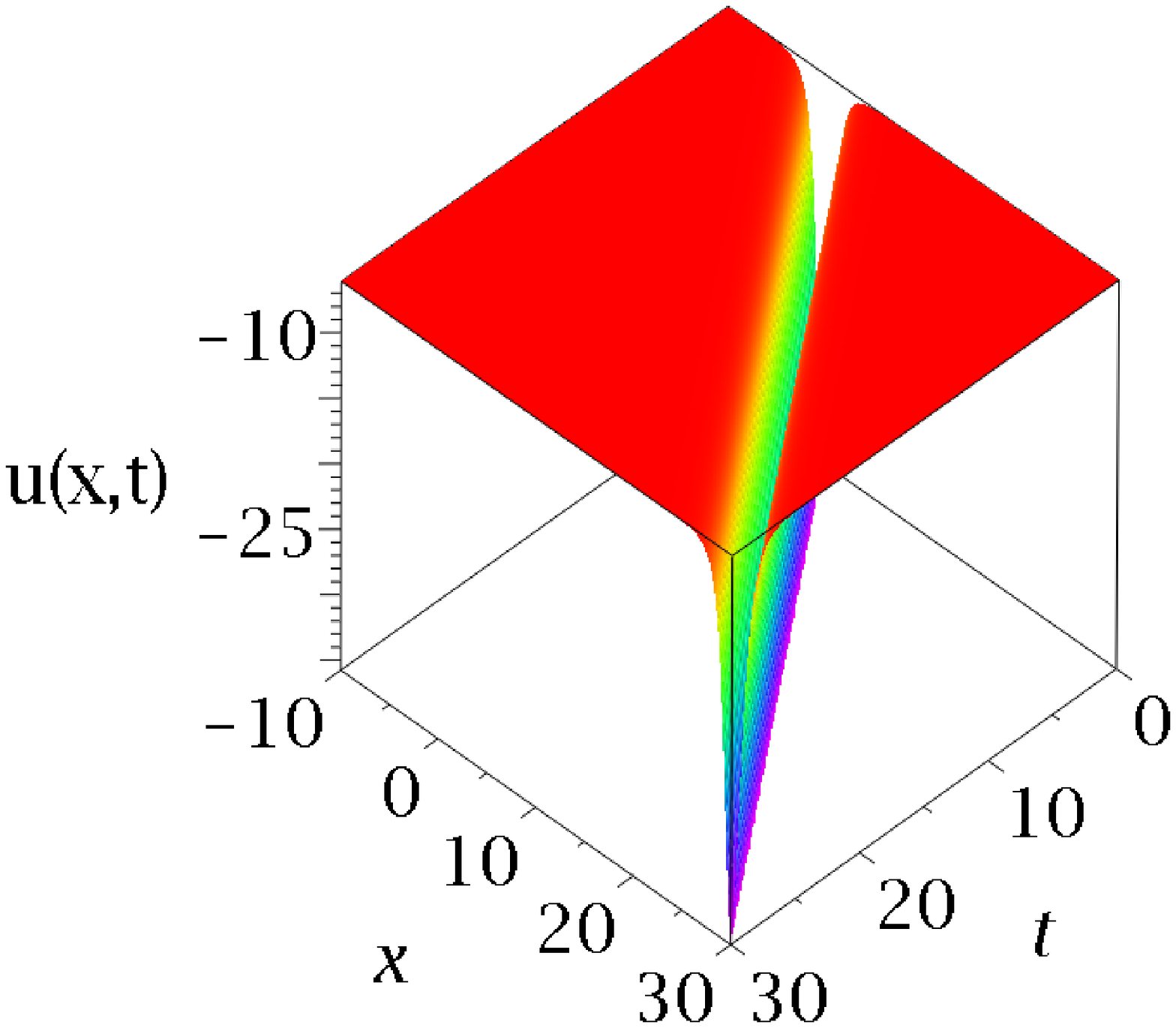}(c)
  \includegraphics[width=5cm, height=5 cm]{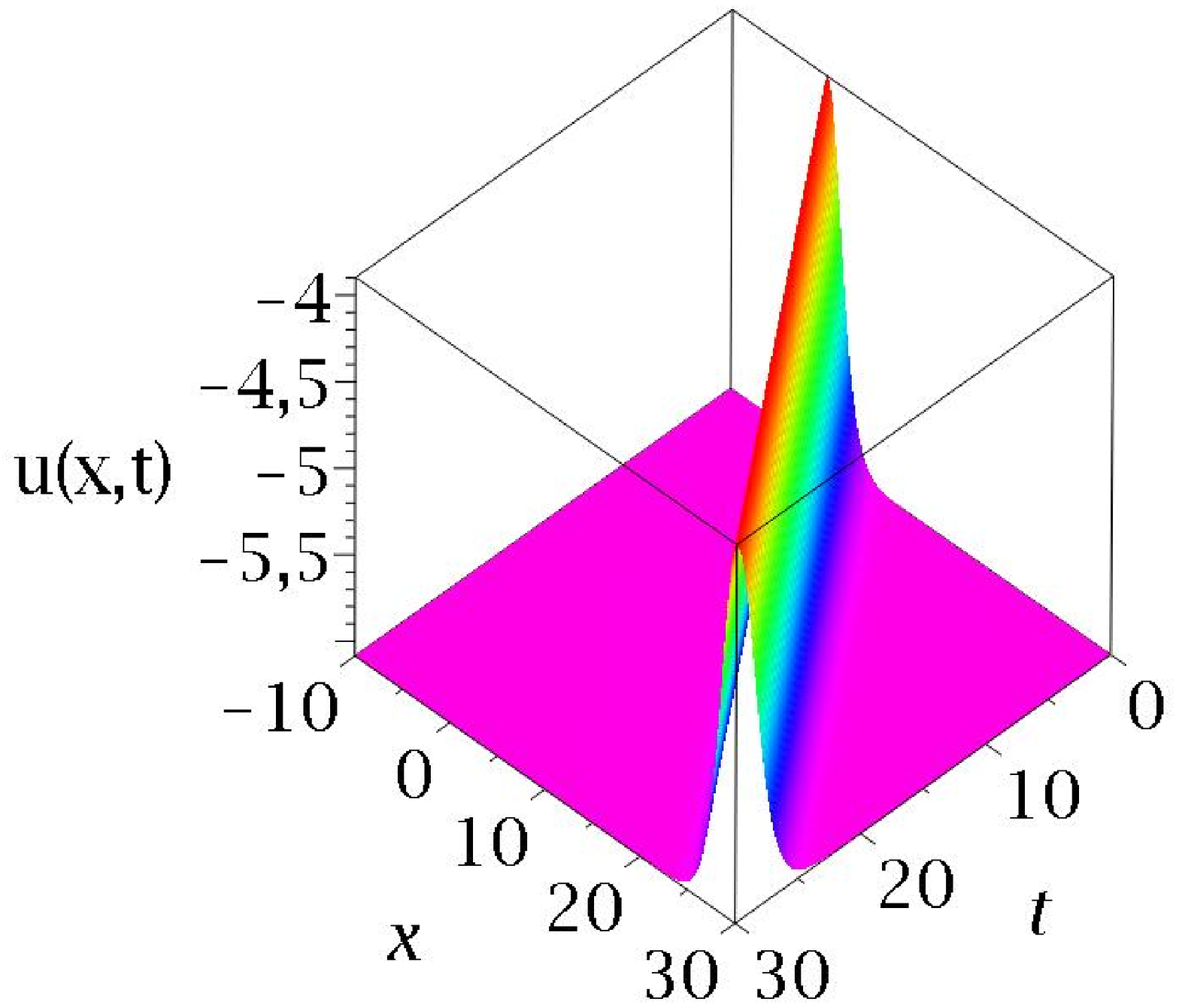}(d)
    \caption{}\label{2}
 \end{center}
\end{figure}

\begin{figure}[!ht]
 \begin{center}
  \includegraphics[width=5cm, height=5 cm]{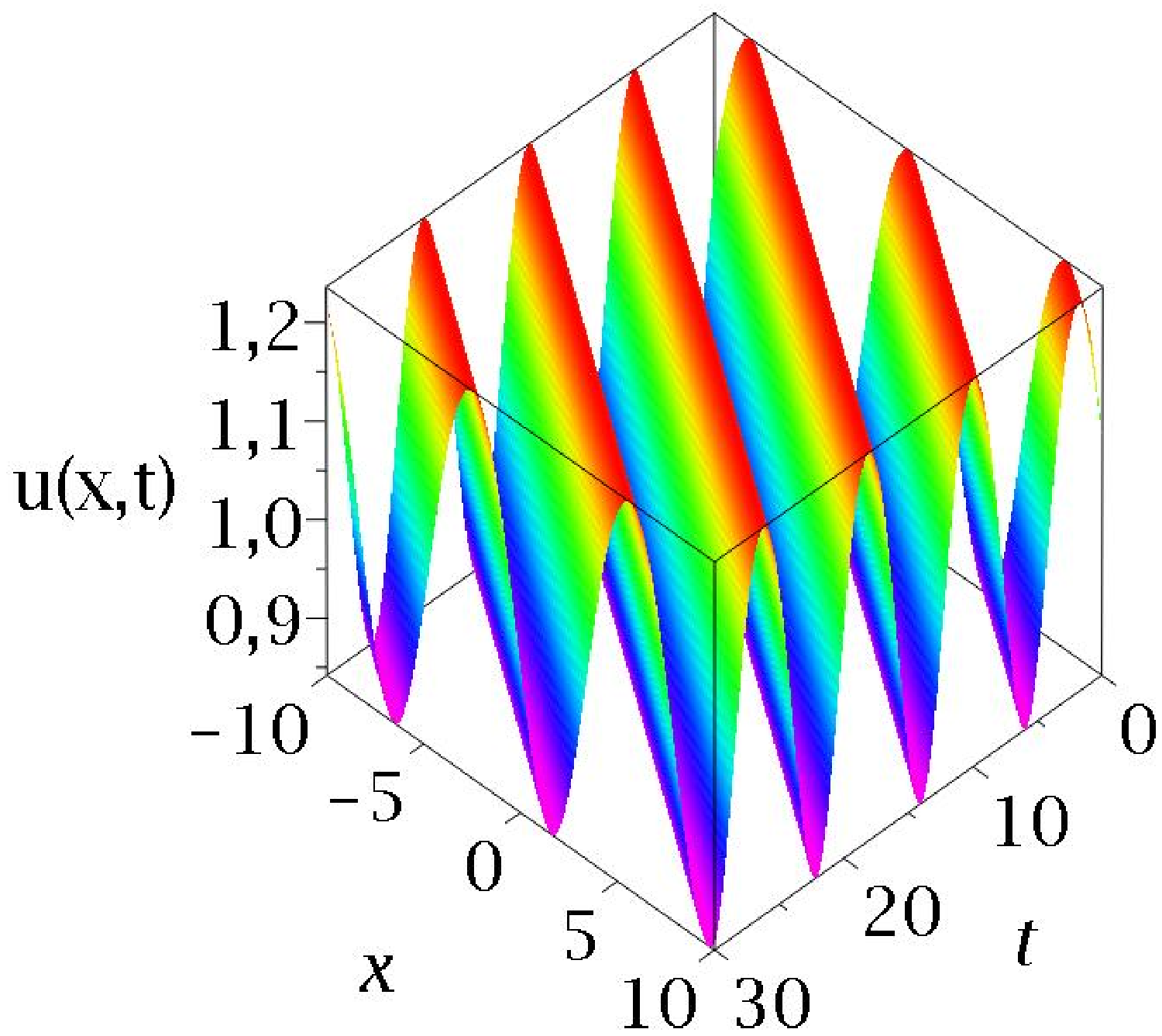}(a)
  \includegraphics[width=5cm, height=5 cm]{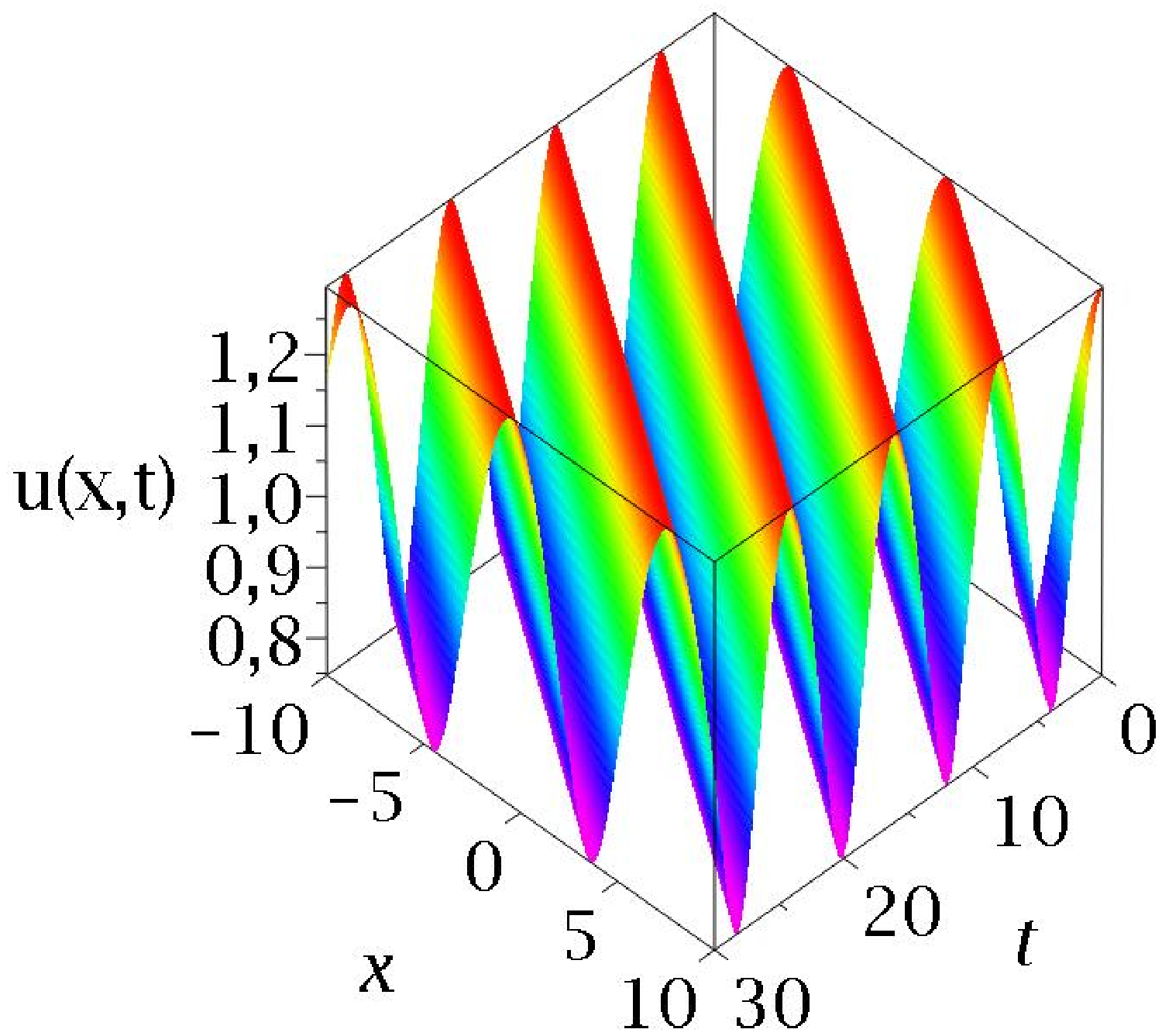}(b)
    \caption{}\label{3}
 \end{center}
\end{figure}
\begin{figure}[!ht]
 \begin{center}
  \includegraphics[width=5cm, height=5 cm]{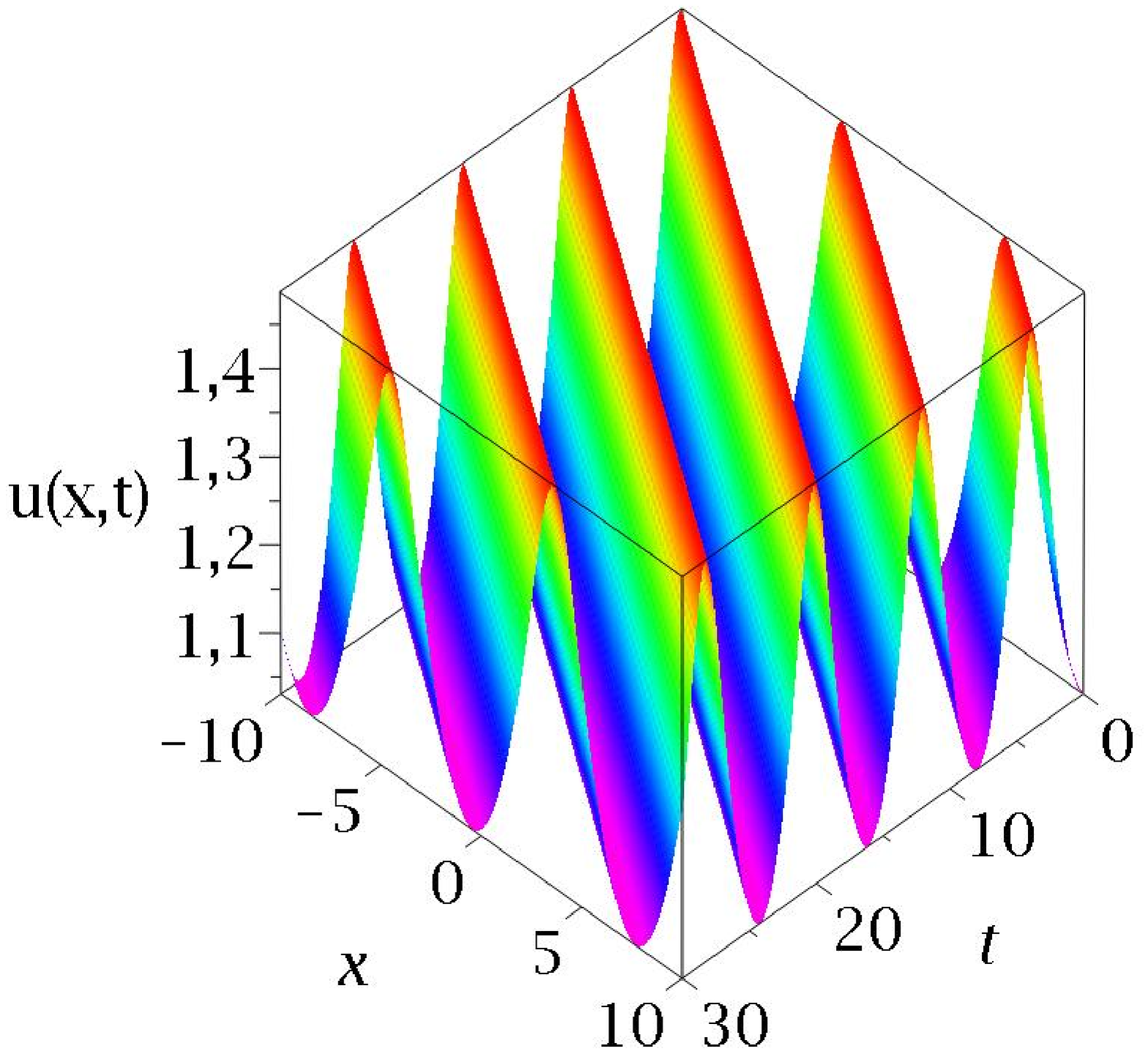}(a)
  \includegraphics[width=5cm, height=5 cm]{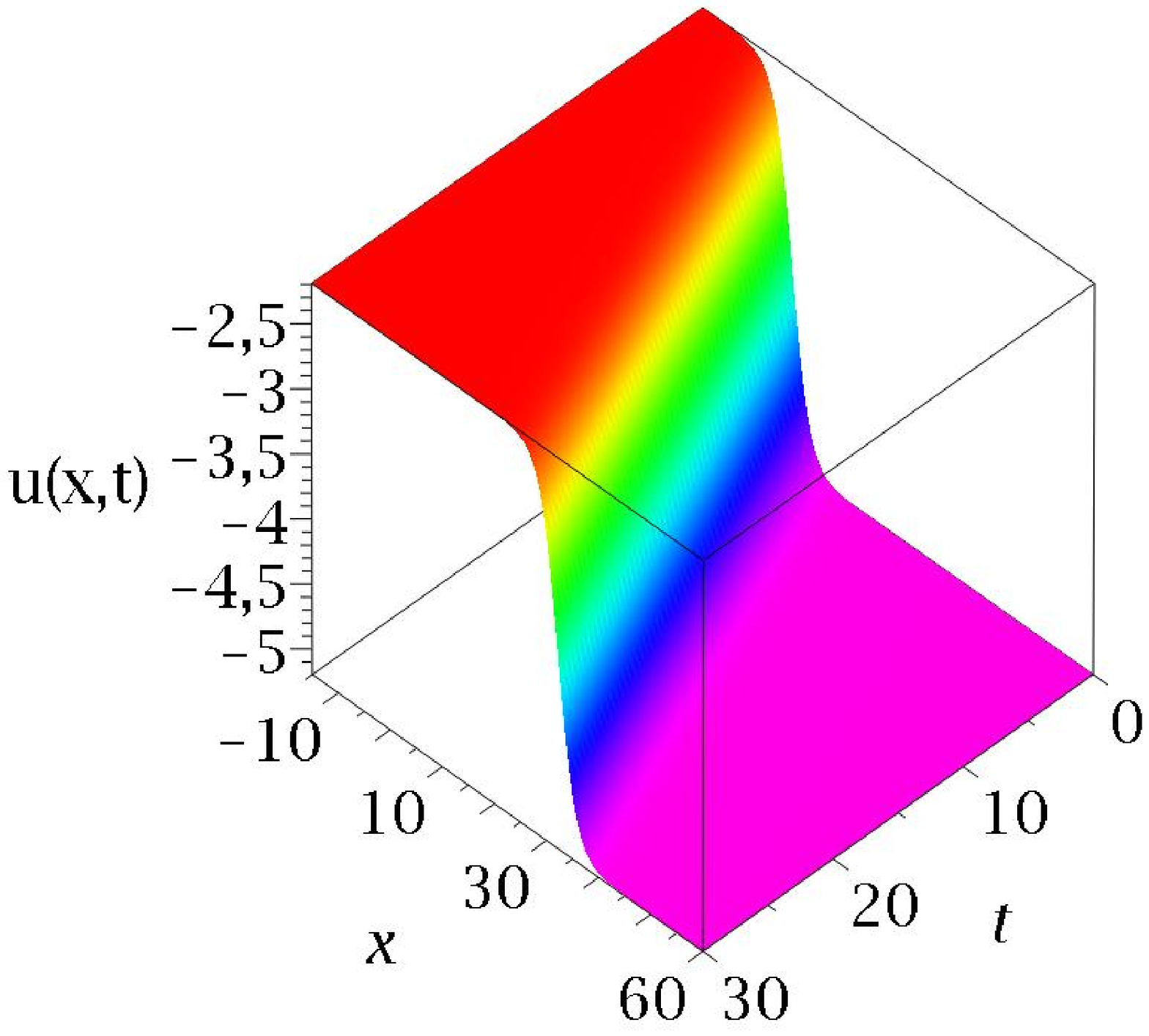}(b)
    \caption{}\label{4}
 \end{center}
\end{figure}
\end{document}